\documentclass[runningheads]{llncs}

\usepackage[np]{numprint}
\usepackage[utf8]{inputenc}
\usepackage[T1]{fontenc}
\usepackage[dvipsnames]{xcolor}
\usepackage{graphicx}
\usepackage{arydshln}
\usepackage{booktabs}
\usepackage{caption}
\usepackage[caption=false]{subfig}
\usepackage{arydshln}
\usepackage{multirow}
\usepackage{import}
\usepackage{amssymb}
\usepackage{fp}
\usepackage{comment}
\usepackage{adjustbox}
\usepackage{balance}
\usepackage[breaklinks]{hyperref}
\usepackage[sort,nocompress]{cite}
\usepackage{cleveref}
\usepackage{xspace}
\usepackage[binary-units]{siunitx}
\DeclareSIUnit{\packet}{p}
\usepackage[nolist]{acronym}

\usepackage{todonotes} %
\presetkeys{todonotes}{fancyline, color=olive!30, inline}{}
\newcommand{\todoog}[1]{}%

\newcommand{\todolv}[1]{}%
\newcommand{\new}[1]{#1}%
\newcommand{\budget}[1]{\todo[color=yellow!40]{\textit{Budget: #1 page(s)}}}
\renewcommand{\budget}[1]{}
\usepackage{pifont} %
\newcommand{\cmark}{\textcolor{PineGreen}{\ding{51}}}%
\newcommand{\xmark}{\textcolor{Red}{\ding{55}}}%
\usepackage[normalem]{ulem}
\usepackage{microtype}

\usepackage{blindtext}

\usepackage{floatrow}
\newfloatcommand{capbtabbox}{table}[][\FBwidth]

\newcommand{\zmap}{ZMap\xspace}
\newcommand{\zmapsix}{ZMapv6\xspace}

\newcommand{\zgrab}{ZGrab2\xspace}
\newcommand{\eg}{e.g.,\xspace}
\newcommand{\ie}{i.e.,\xspace}

\newif\ifcutoptional
\cutoptionalfalse
\DeclareSIUnit{\nothing}{\relax}

\renewcommand*{\paragraph}[1]{%
    \vspace{.5em}
    \noindent
    {\normalfont \bf #1}
}

\usepackage[absolute,showboxes]{textpos}

\begin{document}

\title{Characterizing the VPN Ecosystem in the Wild}

\author{Aniss Maghsoudlou\inst{1} \and
Lukas Vermeulen\inst{1} \and
Ingmar Poese\inst{2} \and
Oliver Gasser\inst{1}}
\institute{Max Planck Institute for Informatics \and
    BENOCS\\
\email{\{aniss,lvermeul,oliver.gasser\}@mpi-inf.mpg.de\\
ipoese@benocs.com}
}

\setlength{\TPHorizModule}{\paperwidth}
\setlength{\TPVertModule}{\paperheight}
\TPMargin{5pt}
\begin{textblock}{0.8}(0.1,0.02)
    \noindent
    \footnotesize
    If you cite this paper, please use the PAM '23 reference:
    Aniss Maghsoudlou, Lukas Vermeulen, Ingmar Poese, Oliver Gasser.
    Characterizing the VPN Ecosystem in the Wild.
    In \textit{Passive and Active Measurement Conference 2023 (PAM '23), March 21--23, 2023.}
    %Springer Nature Switzerland AG 2021, PAM 2021, LNCS 12671, pp. 1--17, 2021.
    %\url{https://doi.org/10.1007/978-3-030-72582-2_32}
\end{textblock}

\maketitle

\setlength{\tabcolsep}{6pt}

\begin{abstract}
\budget{0.5}

With the increase of remote working during and after the COVID-19 pandemic, the use of Virtual Private Networks (VPNs) around the world has nearly doubled.
Therefore, measuring the traffic and security aspects of the VPN ecosystem is more important now than ever. 
VPN users rely on the security of VPN solutions, to protect private and corporate communication.
Thus a good understanding of the security state of VPN servers is crucial.
Moreover, properly detecting and characterizing  VPN traffic remains challenging, since some VPN protocols use the same port number as web traffic and port-based traffic classification will not help.

In this paper, we aim at detecting and characterizing VPN servers in the wild, which facilitates detecting the VPN traffic.
To this end, we perform Internet-wide active measurements to find VPN servers in the wild, and analyze their cryptographic certificates, vulnerabilities, locations, and fingerprints.
We find 9.8M VPN servers distributed around the world using OpenVPN, SSTP, PPTP, and IPsec, and analyze their vulnerability. We find SSTP to be the most vulnerable protocol with more than 90\% of detected servers being vulnerable to TLS downgrade attacks.
Out of all the servers that respond to our VPN probes, 2\% also respond to HTTP probes and therefore are classified as Web servers. 
\new{Finally, we use our list of VPN servers to identify VPN traffic in a large European ISP and} observe that 2.6\% of all traffic is related to these VPN servers.
\end{abstract}
\section{Introduction}

\budget{1.5}

Virtual Private Networks (VPNs) provide secure communication mechanisms, including encryption and tunneling, enabling users to circumvent censorship, to access geo-blocked services, or to securely access an organization's resources remotely.

The COVID-19 pandemic changed Internet traffic dramatically.
Studies investigating the impact of the COVID-19 pandemic on Internet traffic show that streaming traffic being tripled around the world due to remote work, remote learning, and entertainment services \cite{boettger2020,shinan2021,karam2022}.
VPN traffic has been no exception to this major traffic shift.
After the COVID-19 pandemic, the VPN traffic observed in a large European IXP nearly doubled \cite{Feldmann2020LockdownEffect}. 
In a campus network, even a more dramatic increase of 20x has been reported \cite{karam2022}, which shows a prominent growth of remote work and e-learning.
Additionally, several articles find that remote work is here to stay \cite{nytimes2021,forbes2022}.
According to recent statistics from SurfShark \cite{vpn-stats-surfshark-2022}, 31\% of all Internet users use VPNs. 

In order to facilitate network planning and traffic engineering, Internet Service Providers (ISPs) have an interest in understanding the network applications being used by their clients, and how these applications behave in terms of traffic patterns and volume. 
Therefore, detecting and characterizing VPN traffic is an important task for ISPs.
Certain VPN protocols use known port numbers for their operation, e.g. port number 4500 is used for IPsec, and port number 1723 is used for SSTP.
Thus, the traffic using protocols over the known port numbers can easily be detected as VPN traffic.
However, some VPN protocols, e.g. SSTP, and in some occasions, OpenVPN use port number 443 which is commonly used for secure web applications.
This makes it challenging to distinguish between web and VPN traffic. 

Moreover, VPN users might share sensitive private or corporate data over VPN connections.
As the number of cyber attacks has almost doubled after the pandemic \cite{bitaab2020}, it makes Internet users even more aware of their privacy and the security of their VPN connections.
Therefore, investigating the vulnerabilities of the VPN protocols helps to highlight existing shortcomings in VPN security.

Previous studies focused on detecting VPN traffic using machine learning \cite{draper2016, miller2020}, or DNS-based approaches \cite{zain2019,Feldmann2020LockdownEffect}. Some studies have also analyzed the commercial VPN ecosystem \cite{khan2018,ramesh2022}.
However, to the best of our knowledge, this is the first work which conducts active measurements to detect and characterize VPN servers in the wild.

In this paper, we aim to detect, characterize, and analyze the deployment of VPN servers in the Internet using active measurements along with passive VPN traffic analysis. 
Specifically, this work makes the following main contributions:

\begin{itemize}
    \item \textbf{VPN server deployment:} We perform active measurements to the complete IPv4 address space and an IPv6 hitlist for 4 different VPN protocols both in UDP and TCP. We find around 9.8 million IPv4 addresses and 2.2 thousand IPv6 addresses responsive to our probes.
    \item \textbf{VPN security evaluation:} We analyze the detected IP addresses in terms of TLS vulnerabilities, certificates, and geolocation. 
We observe that the United States is the most common location among our detected IP addresses.
We also find that more than 90\% of SSTP servers are vulnerable to a TLS attack and nearly 7\% of the certificates are expired.
    \item \textbf{VPN traffic analysis:} We analyze passive traffic traces from a large European ISP, we find that 2.6\% of the traffic uses our list of VPN servers as either source or destination address.
Moreover, we use rDNS data along with DNS records from a large European ISP to compare our results with previous work looking into VPN classification \cite{Feldmann2020LockdownEffect}. We find that using our methodology, we find 4 times more VPN servers in the wild.
    \item \textbf{VPN probing tool: } We develop new modules for \zgrab \cite{zgrab} to send customized VPN probes. We make these modules publicly available \cite{zgrab2-vpn} to foster further research in the VPN ecosystem.
\end{itemize}

\section{Background}
\budget{2}

VPNs establish cryptographically secured tunnels between different networks and can be used to connect private networks over the public network. Thus, a proper VPN connection should be encrypted in order to prevent eavesdropping and tampering of VPN traffic. The tunneling mechanism of a VPN connection also provides privacy since the traffic is encapsulated. Therefore, users remotely accessing a private network appear to be directly connected.

While the exact tunneling process varies depending on the underlying VPN protocol, it is quite common to categorize VPNs in two different groups:

\begin{itemize}
    \item \textbf{Site-to-site VPNs}: In this configuration, a VPN is used to connect two or more networks of geographically distinct sites. This is common for companies with branches in different locations.
    \item \textbf{Remote access VPNs}: This kind of VPN connection is mainly used by individual end-users in order to connect to a private network. 
\end{itemize}

\subsection{VPN Usage}

The usage of VPNs has evolved over the past three decades. David Crawshaw \cite{Crawshaw2020} gives a very comprehensive overview of how and why VPNs changed over the years. While in the earlier days of the Internet, they were primarily used by companies to connect their geographically distinct offices, VPNs nowadays provide a variety of use cases for individuals as well and are used by millions of end-users around the globe. Use cases include:

\begin{itemize}
    \item \textbf{Privacy preservation}: The encrypted VPN tunnels provide end-users the means to preserve their privacy.
    \item \textbf{Censorship circumvention/accessing geo-blocked content}: Specific services might be censored in some countries or geographically restricted. By connecting to a VPN server in a different country, it is still possible to access such content since it would now appear as if the user was located in a different country.
    \item \textbf{Remote access}: It is common to use VPNs to remotely access restricted resources or to connect with an organization's network. This usage scenario has gained importance especially during the COVID-19 pandemic among employees and students alike due to remote working.
\end{itemize}

Different usage patterns, the general understanding of the functionality of VPNs, and awareness of potential risks vary between different demographic groups. Dutkowska-Zuk et al. \cite{Dutkowska-Zuk2022} studied how and why people from different demographic backgrounds use VPN software primarily comparing the general population with students. They found that the general population is more likely to rely on free, commercial VPN solutions to protect their privacy. Students, on the other hand, rather resort to VPN software for remote access or to circumvent censorship and access geographically blocked services with an increased use of institutional VPNs. Generally, they found that, while most VPN users are concerned about their privacy, they are less concerned about data collection by VPN companies.  

Especially during the COVID-19 pandemic, VPNs increasingly gained significance. The pandemic and the resulting lockdowns caused many employees and students to work and study remotely from home. Feldmann et al. \cite{Feldmann2020LockdownEffect} analyzed the effect of the lockdowns on the Internet traffic. Their work included the analysis of how VPN traffic shifted during the pandemic. They detected a traffic increase of over 200\% for VPN servers identified based on their domain with increased traffic even after the first lockdowns. These findings highlight the rising significance of VPNs. With progressing digitalization, VPN traffic can be expected to increase even further.

\subsection{VPN Protocols}
We want to cover as many protocols as possible including some of the most prominent ones like OpenVPN and IPsec. The functionality of a VPN connection establishment varies depending on the underlying VPN protocol. Table \ref{table:protocol_overview} gives an overview of all the VPN protocols we consider with general information on their underlying protocols. Among them, especially PPTP, which was the first actual VPN protocol standardized in 1999 (see RFC 2637 \cite{rfc2637}), can be considered rather outdated and it is not recommended to be used anymore \cite{pptp-security-advisory,Crawshaw2020}.

WireGuard is the most \new{modern} protocol at the moment. It is much more simplistic than, \eg OpenVPN or IPsec and incorporates state-of-the-art cryptographic principles. 

\begin{table}
\begin{adjustbox}{max width=\textwidth}
        \begin{tabular}{lllll}
        \toprule
        VPN protocol & Transport protocol & Port      & (D)TLS-based    & \new{Server} detection possible    \\ \midrule
        IPsec/L2TP   & UDP                & 500       & \xmark          & \cmark                        \\ 
        OpenVPN      & UDP \& TCP         & 1194, 443 & \cmark          & partially                     \\ 
        SSTP         & TCP                & 443       & \cmark          & \cmark                        \\ 
        PPTP         & TCP                & 1723      & \xmark          & \cmark                        \\ 
        AnyConnect   & UDP \& TCP         & 443       & \cmark          & \xmark                        \\ 
        WireGuard    & UDP                & 51820     & \xmark          & \xmark                        \\ \bottomrule
        \end{tabular}
    \end{adjustbox}
    \caption{Overview of VPN protocols showing the transport protocol, port, (D)TLS encryption, and possible detection.}
    \label{table:protocol_overview}
\end{table}

\section{Methodology}
\budget{2}

In this section, we introduce our methodology for our passive and active measurements. We perform Internet-wide measurements in order to detect VPN servers in the wild and create hit lists of identified VPN servers. Based on those results, we conduct follow-up measurements to fingerprint the VPN servers and further analyze them in terms of security. 
Finally, we look for the detected IP addresses in the traffic from a large European ISP to find out the amount of VPN traffic.
\subsection{VPN Server Detection}\label{subsection:vpn-server-detection}

Our measurements to detect VPN servers include the whole IPv4 address range as well as over 530 million non-aliased IPv6 addresses from the \textit{IPv6 Hitlist Service} \cite{gasser2018clusters,zirngibl2022rusty}. We send out the connection initiation requests that are used in the connection establishments of the different VPN protocols. For UDP-based protocols, we use \zmap \cite{zmap} (ZMapv6 for IPv6\cite{zmapv6}), a transport-layer network scanner, to directly send out UDP probes. If the VPN protocol is TCP-based, we first use \zmap to find targets with the respective open TCP ports using TCP SYN-scans. We then use \zgrab \cite{zgrab} to send out the actual VPN requests. \zgrab works on the application layer. It can be used complementary to \zmap for more involved scans. It also allows us to implement custom modules needed for our VPN requests over TCP and TLS. 

We identify an address as a VPN server based on the responses we receive to our initiation requests. If the parsed response satisfies the format of the expected VPN response, the target is classified as a VPN server. To completely detect the VPN ecosystem for a specific server, we might have to take several server configurations into account and perform multiple measurements for a single protocol accordingly. Apart from that, for some protocols and configurations, we require knowledge of cryptographic key material which we do not have since we perform measurements in the wild. Therefore, we cannot detect the entirety of the VPN ecosystem with our method. The last column of \Cref{table:protocol_overview} summarizes for which protocol we are able to detect VPN servers. When OpenVPN servers specify the so-called \textit{tls-auth} directive, an HMAC signature is required in all control messages. This means that we can only craft requests without HMACs and hence detect only a subset of all OpenVPN servers.

As mentioned above, for some protocols, it might also be necessary to consider different configurations. For IPsec, \eg we suggest seven different cipher suites in the initiation request. Apart from that, we have to specify a key exchange method in the OpenVPN requests. Out of the two possible key exchange methods, namely \textit{key method 1} and \textit{key method 2}, key method 1 is considered insecure and is therefore deprecated \cite{key-method-1}. We therefore specify key method 2 in our initiation requests and then perform a follow-up scan where we suggest the deprecated key exchange method to identified OpenVPN servers to investigate how many of them might still support key method 1.

\subsection{TLS Analysis}
For the TLS-based VPN protocols, which include SSTP and OpenVPN over TCP, we perform follow-up measurements to further fingerprint the servers and assess them in terms of security. For that, we collect TLS certificates of the VPN servers to analyze them for expiry, check for self-signed certificates and investigate how many of them are snake oil certificates. We characterize a certificate as a snake oil certificate if the common name (CN) of the subject and issuer are both specified as \textit{localhost} or \textit{user.local}. For the certificates signed by a Certificate Authority (CA), we collect the most common issuing organizations. We gather domain names corresponding to the responsive IP addresses using reverse DNS (rDNS) look-ups, and collect certificates with and without the Server Name Indication (SNI) extension using these domain names and compare them against each other. SNI can be used by the client in the TLS handshake in order to specify a hostname for which a connection should be established. This might be necessary in cases where multiple domain names are hosted on a single address. Finally, we test if the servers are susceptible to the \textit{Heartbleed} \cite{heartbleed} vulnerability as well as a series of TLS downgrade attacks. The Heartbleed attack is based on the Heartbeat Extension \cite{rfc6520} of the OpenSSL library. In TLS downgrade attacks, we try to force a server to establish a connection using an outdated SSL/TLS version or using insecure cipher suites by suggesting those outdated primitives in the TLS handshake. \Cref{tab:tls-vulnerabilities} summarizes all the vulnerabilities and their requirements, \ie what we have to test for or the version or cipher suite to which we try to downgrade the TLS connection. For instance, in order to check if a server is vulnerable to the FREAK attack, we suggest any SSL/TLS version and only RSA\_EXPORT cipher suites in the TLS handshake.

\subsection{Fingerprinting}
We try to infer more information on the VPN servers based on our connection initiation requests as well as from follow-up measurements in order to further categorize them. 

One aspect we examine is the server software deployment. For SSTP and PPTP, we can extract information on the software vendor directly from the responses to our initiation requests.

Furthermore, we perform OS detection measurements on a subset of 1000  VPN servers for each protocol using \textit{Nmap} \cite{nmap}, a network scanner that can be used for network discovery among other things.
We use Nmap's \textit{fast} option and target 100 instead of 1000 ports to decrease runtime and parse the results for the most common open ports and OS guesses. With those results, we can learn more about the VPN server infrastructure and potential other services running on the same servers.

\subsection{VPN Traffic Analysis}
The active measurement in \Cref{subsection:vpn-server-detection} provides us with a list of IP addresses, namely VPN hitlist,  which are responsive to at least one VPN protocol initiation request. 
We look for these IP addresses in the DNS records gathered from the DNS resolvers at a large European ISP during a 1-hour period to learn about the domain names these IP addresses are associated with. 
We do not expect to find all the detected IP addresses in these DNS responses. Therefore, for any remaining IP address, we use reverse DNS resolution to find the corresponding domain names.

Then, we look for the IP addresses from our VPN hitlist on over a week of network flow data from the ISP to find out the amount of traffic associated with the VPN hitlist and compare the results with a port-based VPN traffic detection, and also a state-of-the-art approach. 

\subsection{\new{Ethical Considerations}}
\noindent\new{\textbf{Active scanning.} 
We follow best current practices \cite{kenneally2012menlo,partridge2016ethical} to avoid potential harm to the networks we scan.
We make sure that our prober IP address has a meaningful DNS PTR record pointing to our Web server which allows for requesting an opt-out from being scanned.
We also limit our scanning rate and perform probing in a randomized order.
We plan to notify the VPN providers about their servers' vulnerabilities.}

\noindent\new{\textbf{ISP data.} 
All the data related to the ISP is processed on the ISP's premises. 
We do not copy, transfer, or store any data outside the dedicated servers that the ISP uses for its NetFlow analysis.}

\section{Active Measurements of the VPN Server Ecosystem}
In this section, we go through the results from our Internet-wide active measurement using different VPN protocols. 
We discuss the characteristics of the responsive servers such as geographical locations, VPN protocols, etc.
Then, we analyze their vulnerabilities and try to fingerprint them based on the gathered information. 
\begin{figure}
  \begin{center}
  \includegraphics[width=\linewidth]{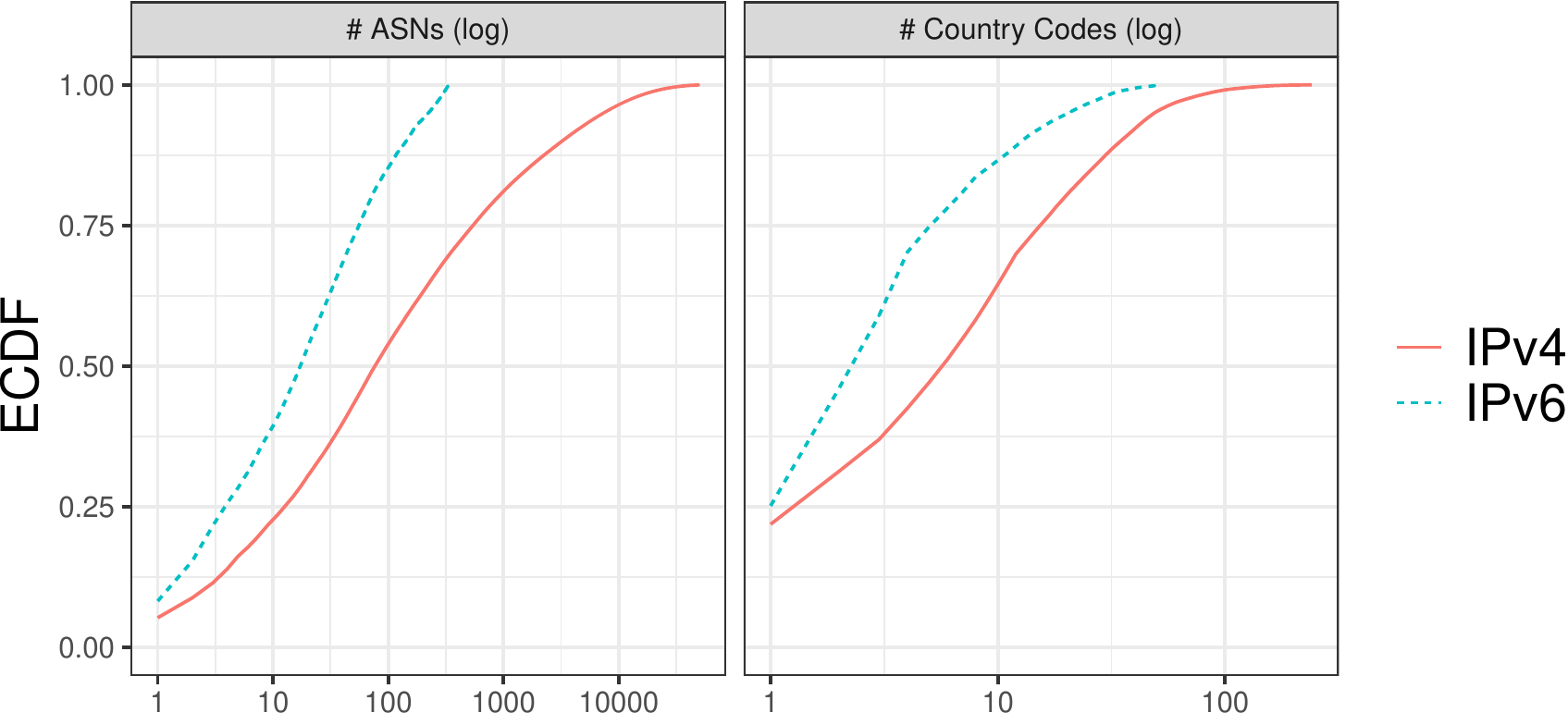}
  \end{center}
  \caption{\new{Cumulative distribution of number of ASes (left) and number of countries (right) corresponding to the responsive IPs.}}
  \label{fig:cdf-as-cc}
\end{figure}

\subsection{Responsive Servers}
\budget{2.5}

In total, we find 9,817,450 responsive IPv4 addresses with our probes that we can identify as VPN servers.

\noindent\new{\noindent \textbf{rDNS.} We investigate the reverse DNS records corresponding to the responsive IPv4 addresses.
We aggregate results on the second-level domain and sort them based on the number of responsive IPs that they correspond to.
We find that all the top 10 domain names belong to telecommunication companies (\eg Open Computer Network, a large Japanese ISP, and Telstra, an Australian telecommunications company).
Next, we filter all rDNS records which contain \textit{vpn} in their second-level domain names in order to detect commercial VPN providers.
We find a single domain related to PacketHub which manages IP addresses for several companies, including NordVPN, a major commercial VPN provider. 
This domain name ranks 60th among all rDNS second level domains.}

\noindent\textbf{AS analysis}.
\new{\Cref{fig:cdf-as-cc} shows the distribution of ASes to which our responsive IP addresses belong.} The responsive IP addresses are originated by 49625 and \new{334} ASes in total, while top 10 ASes contribute to 22\% and 38\% of the IP addresses, for IPv4 and IPv6 respectively, as shown in \Cref{fig:cdf-as-cc}. Top 10 ASes for IPv4 responsive addresses are all telecommunication companies, while out of the top 10 ASes for IPv6 responsive addresses, 8 are telecommunication companies and 2 are academy-related ASes.
\new{\Cref{tab:asn-ipv4,tab:asn-ipv6} further summarize the top 10 AS numbers as well as the AS names or organizations and the number of VPN servers that are registered within the respective AS.
As can be seen, most top ASes are large ISP networks.}

\new{Moreover, we investigate the top ASes for commercial VPN providers.
As shown by Ramesh et al. \cite{ramesh2022} it is quite common for commercial VPN providers to use shared infrastructure. 27 providers, including popular companies such as NordVPN, Norton Secure VPN, or Mozilla VPN, use the same AS, namely AS 9009 operated by M247 Ltd.
This AS is also visible in our measurements and it ranks 14th with 74,894 identified VPN servers (0.76\% of all addresses). Furthermore, Ramesh et al. \cite{ramesh2022} find that some IP blocks in AS 16509 (Amazon) are shared across Norton Secure VPN and SurfEasy VPN. AS 16509 lands on rank 20 of our list being shared by almost 60,000 VPN servers (0.6\% of all addresses). Another AS known to be used by VPN providers is AS 60068---again operated by M247 Ltd.--- which is used by NordVPN and CyberGhost VPN. It ranks on place 178 of our list with 6,898 VPN servers (0.07\% of all addresses).
Overall, we find that although the top ASes are dominated by large ISPs, a considerable number of VPN servers are located in ASes used by commercial VPN providers.}

\begin{table}[]
    \RawFloats
    \parbox{.45\linewidth}{
        \centering
        \resizebox{.45\textwidth}{!}{
        \begin{tabular}{@{}rlr@{}}
            \toprule
            AS number & AS name & \multicolumn{1}{l}{VPN servers} \\ \midrule
            4134      & ChinaNet    & 515,830      \\
            7922      & Comcast     & 356,327      \\
            1221      & Telstra     & 257,821           \\
            3320      & Deutsche Telekom   & 242,433      \\
            4766      & Korea Telecom        & 228,863      \\
            4713      & NTT Communications      & 145,286      \\
            7018      & AT\&T         & 137,698      \\
            4837      & China Unicom        & 133,861      \\
            3462      & HiNet  & 119,612      \\
            20115     & Charter Communications       & 97,109       \\ \bottomrule
        \end{tabular}}
        \caption{IPv4: AS numbers, AS names and number of VPN servers belonging to the ASes. \label{tab:asn-ipv4}}
    }
    \hfill
    \parbox{.45\linewidth}{
        \centering
        \resizebox{.45\textwidth}{!}{
        \begin{tabular}{@{}rlr@{}}
            \toprule
            AS number & AS name & \multicolumn{1}{l}{VPN servers} \\ \midrule
            7922      & Comcast                  & 183 \\
            63949     & Akamai     & 159 \\
            12322     & Proxad Free SAS               & 138 \\
            7506      & GMO Internet Group     & 89  \\
            9009      & M247 Ltd                      & 63  \\
            9370      & Sakura Internet Inc & 58  \\
            14061     & DigitalOcean              & 55  \\
            2516      & KDDI Corporation              & 54  \\
            7684      & Sakura Internet Inc & 39  \\
            680       & DFN-Verein   & 36  \\ \bottomrule
        \end{tabular}}
        \caption{IPv6: AS numbers, AS names and number of VPN servers belonging to the ASes. \label{tab:asn-ipv6}}}
\end{table}

\begin{figure}[t!]
  \begin{center}
  \includegraphics[width=\linewidth]{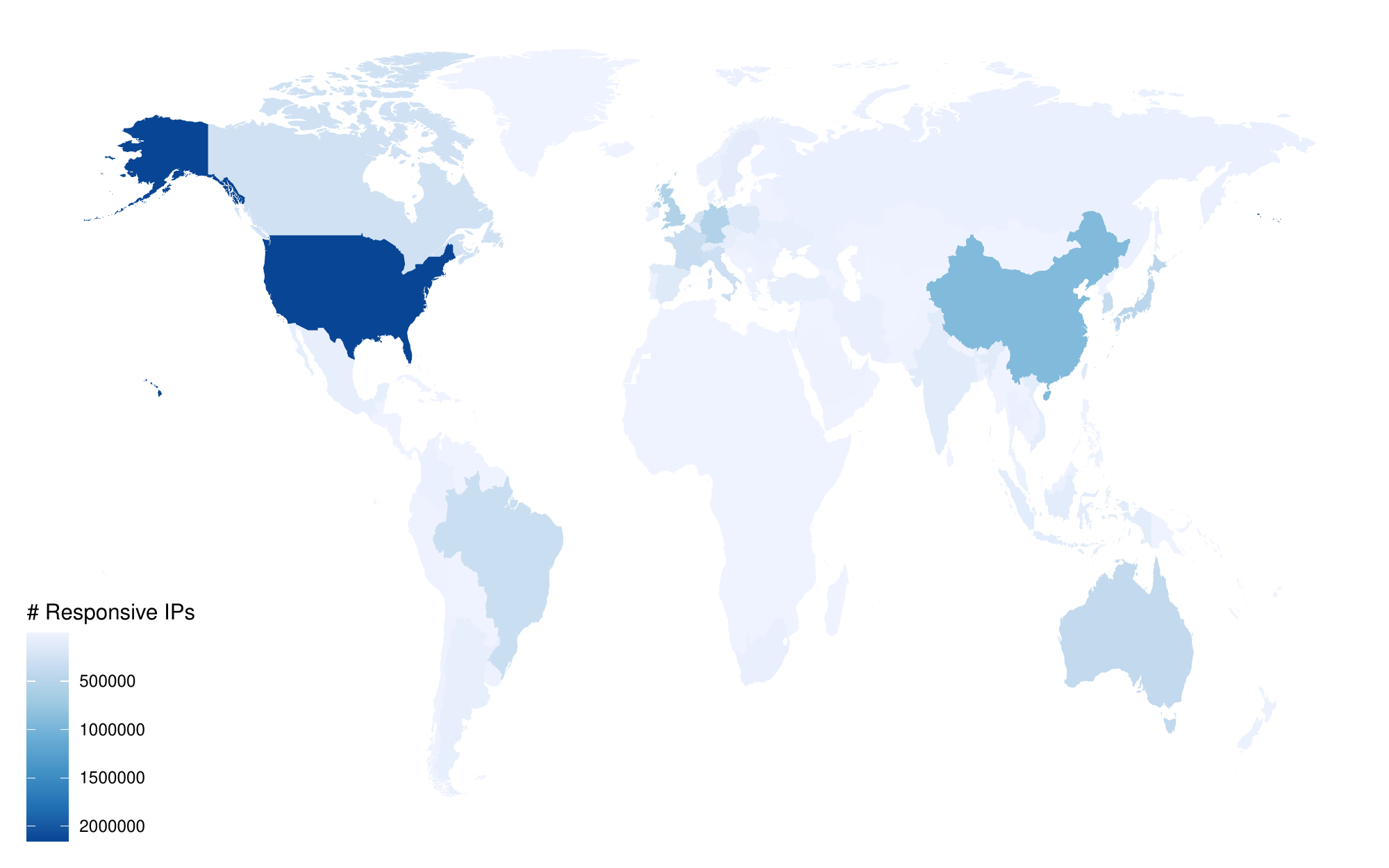}
  \end{center}
  \caption{Geographical distribution of responsive IPv4 addresses per country.}
  \label{fig:heatmap-countries}
\end{figure}

\noindent\textbf{Geolocation}.
We use Geolite Country Database \cite{geolite} to determine the location of the responsive IP addresses. 
\Cref{fig:heatmap-countries} shows a heatmap of the number of responsive IPv4 addresses per country.
We observe that responsive IP addresses are scattered all over the world, in total over 241 and 52 countries for IPv4 and IPv6, respectively. 
However, 64\% and 86\% of IP addresses belong to the top 10 countries for IPv4 and IPv6 respectively. 
Top 3 countries contributing to IPv4 responsive addresses are the United States, China, and UK, while top 3 countries for IPv6 are the United States, Japan, and Germany. 

\begin{table}
    \centering
    \begin{tabular}{lr}
    \toprule
        VPN protocol &  Detected servers \\ \midrule
        IPsec &  7,008,298 \\ 
        PPTP & 2,424,317 \\ 
        OpenVPN &  1,436,667 \\ 
        SSTP &  187,214 \\ \bottomrule
    \end{tabular}
    \caption{Number of detected VPN servers per protocol.}
    \label{tab:vpn-protocols}
\end{table}

\subsection{VPN Protocols}\label{active:vpn-protocols}
\budget{2}

We are able to detect servers for IPsec, PPTP, OpenVPN without tls-auth, and SSTP. \Cref{tab:vpn-protocols} summarizes our findings. Our IPsec UDP probes yield by far the most responsive VPN servers. It might seem surprising that we find such a large number of PPTP servers in contrast to OpenVPN and SSTP considering that PPTP is far more outdated and OpenVPN is one of the most prominent VPN protocols. However, we have to keep in mind that we can only detect a subset of the whole OpenVPN ecosystem since some configurations require knowledge of cryptographic key material \new{as explained in \Cref{subsection:vpn-server-detection}}. Apart from that, SSTP can only be used for remote access connections, whereas PPTP used to be the most widely deployed VPN protocol. We can assume that a large number of the detected PPTP servers are quite outdated, yet still running. 

\begin{figure}[t!]
        \begin{center}
        \includegraphics[width=0.7\linewidth]{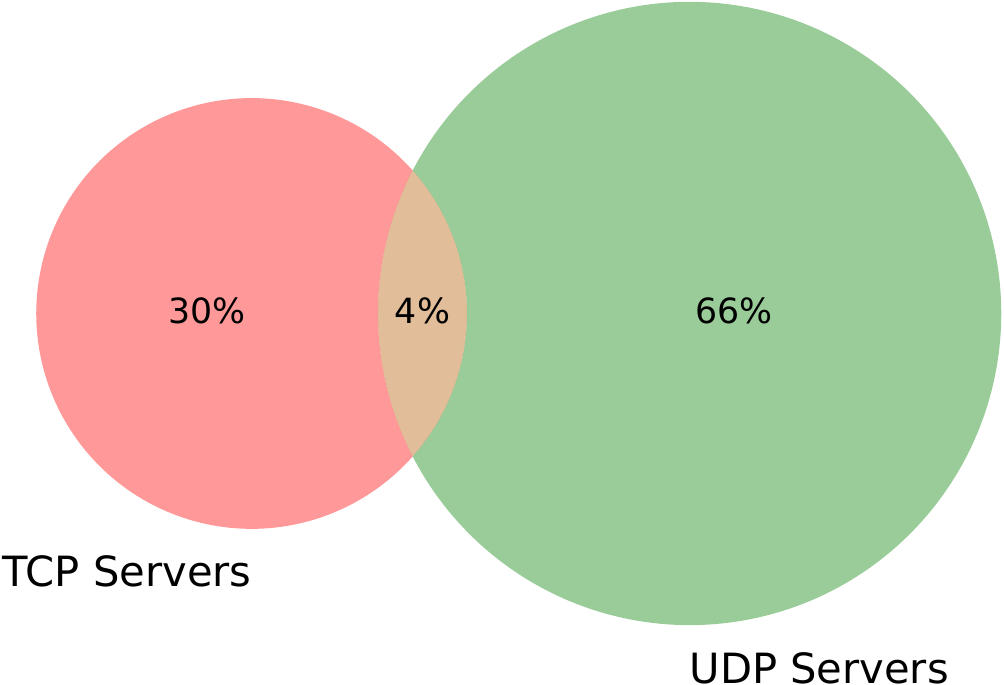}
        \end{center}
        \caption{Intersection of OpenVPN UDP and TCP servers.}
        \label{fig:venn-openvpn}
      \end{figure}

Out of the around 1.4 million OpenVPN servers, 1,011,178 were detected over UDP and 482,956 over TCP. Considering that the TCP version of OpenVPN is generally rather considered as a fallback option, this disparity is to be expected. \Cref{fig:venn-openvpn} visualizes the intersection of those two address sets in a Venn diagram. We can see that the majority of the servers supports only a single transport protocol.
    
    \begin{figure}
        \begin{center}
        \includegraphics[width=\linewidth]{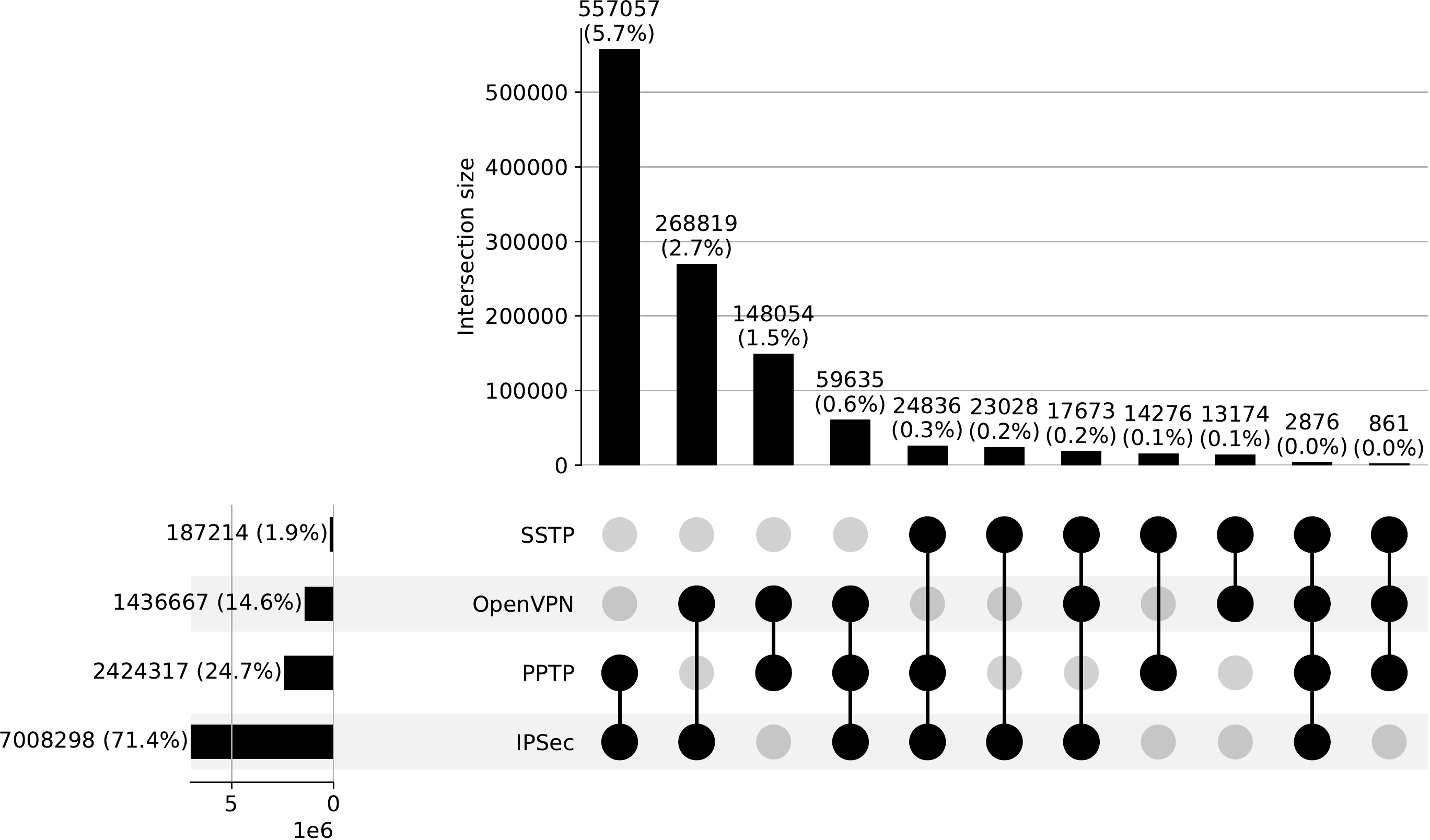}
        \end{center}
        \caption{VPN protocol summary: Number of detected VPN servers for each protocol and the intersection between all protocols.}
        \label{fig:upset-plot}
      \end{figure}

\noindent\textbf{Overlap between protocols}.
In the next step, we compare the IP address sets for the four protocols to depict their intersections and to find out how many of the servers support more than one VPN protocol. \Cref{fig:upset-plot} summarizes those findings in an upset plot. The horizontal bars on the left visualize the sizes of the four protocol sets. The vertical bars represent the different intersections and the sets to be considered are indicated by the black dots below the vertical bars. The first bar on the left, \eg represents the number of VPN servers supporting both PPTP and IPsec with roughly 550,000 servers making up for around 5.7\% of the whole detected VPN \new{server} ecosystem. The second bar on the right, on the other hand, represents the number of servers supporting all four protocols, which is close to zero with only around 2.8 thousand servers.

We can see that the majority of all VPN servers \new{support} only \new{one of the four protocols we consider in this work}. Since commercial VPN providers usually offer a variety of different VPN protocols to choose from, it is possible that a large percentage of the servers supporting several protocols are commercial. This might be the case especially for the ones supporting three or four protocols. \new{We investigate the rDNS records corresponding to the servers supporting all the four protocols, and find that there are no commercial VPN provider in the top 10 second-level domains.
All in all, we find that commercial VPN providers account for only a fraction of the entire VPN \new{server} ecosystem considering the supported protocols.}
    
\noindent\textbf{Different protocol versions}.
Some VPN protocols might include different versions or configurations, like OpenVPN, for instance. We therefore try to trigger VPN responses from OpenVPN servers suggesting the outdated key exchange method key method 1. We also try to trigger responses with random HMAC signatures. 

We find that only 84 of the roughly 1.4 million servers accept our random signature. Apart from that, none of the detected servers support the insecure key exchange method. While most of the servers ignored our requests, we still received around 6,500 responses specifying the default key exchange method key method 2. We can therefore conclude that key method 1 is truly deprecated in the OpenVPN ecosystem.

\subsection{Security Analysis}
\budget{3}

\noindent\textbf{TLS certificate analysis}.
We collect TLS certificates for the TLS-based VPN servers which include SSTP and OpenVPN over TCP and consider only unique certificates. For that, we compare the certificate fingerprints, \ie the unique identifier of the certificate, to make sure we do not consider the same certificate more than once. Some certificates, however, do not include a fingerprint. Therefore, the number of certificates that we analyze in the end might be higher than the number of unique certificates. For OpenVPN, we find 129,143 unique certificates with a fingerprint for 312,095 servers. The most frequently occurring certificate is collected over 10,000 times and is issued for \textit{www.update.microsoft.com}. For SSTP, there are 104,988 fingerprints for 184,047 servers. We detect a certificate issued for \textit{*.vpnauction.com} 2561 times and one for \textit{*.trust.zone} 1194 times. These are commercial VPN providers that seem to use the same certificate for all of their VPN servers. While we are able to collect certificates for nearly all of the SSTP servers, we only receive TLS certificates for around 65\% of the detected OpenVPN servers. This is most likely caused by the fact that OpenVPN performs a variation of the standard TLS handshake during connection establishment. Therefore, some of the servers might not respond when trying to initiate a regular TLS handshake.

    \begin{table}
        \begin{tabular}{@{}lll@{}}
        \toprule
                               & OpenVPN TCP & SSTP    \\ \midrule
        Expired                & 6080 (3.8\%)       & 13,370 (9\%)  \\ 
        Self-issued            & 109,965 (69\%)     & 39,889 (28\%)  \\ 
        Self-signed            & 109,825 (69\%)     & 34,725 (24\%)  \\
        \midrule
        All certificates & 158,705     & 143,517 \\ 
        \bottomrule
        \end{tabular}
        \caption{Expired, self-issued, and self-signed TLS certificates for OpenVPN and SSTP.}
        \label{tab:tls-certificates}
    \end{table}

    \Cref{tab:tls-certificates} summarizes the results of the certificate analysis and contains the number of certificates that we analyzed after filtering out unique certificates and certificates without fingerprints. We detect a large number of self-issued or self-signed certificates for both protocols. Out of the self-issued certificates, we characterize only around 4.7\% as snake oil certificates for SSTP and close to zero for OpenVPN with around 0.4\%. However, 33\% of the self-issued SSTP certificates contain \textit{softether}, an open-source and multi-protocol VPN software, in the CN fields. 13\% specify an IPv4 address in the CN sections. Upon looking at the organization field, we find over 21,000 different organizations where almost 14,000 specify no organization at all. For the OpenVPN certificates, we find that around 77\% of the self-issued certificates include the \textit{Fireware web CA} as CNs specifying \textit{WatchGuard} as organization. For the rest, we detect more than 21,000 different organizations.

    \begin{table}
        \RawFloats
        \parbox{.45\linewidth}{
            \centering
            \begin{tabular}{@{}ll@{}}
                \toprule
                Issuer          & Certificates \\ \midrule
                Stormshield           & 8966   \\ 
                Let's Encrypt         & 7463   \\ 
                Sectigo Limited       & 2610   \\ 
                Digicert Inc.         & 1897   \\ 
                GoDaddy.com, Inc.     & 1332    \\ \bottomrule
                \end{tabular}
            \caption{Certificate issuer distribution for OpenVPN servers. \label{tab:organizations-openvpn}}
        }
        \hfill
        \parbox{.45\linewidth}{
            \centering
            \begin{tabular}{@{}ll@{}}
                \toprule
                Issuer          & Certificates \\ \midrule
                DigiCert, Inc.    & 45,156   \\ 
                Sectigo Limited   & 12,795   \\ 
                GoDaddy.com, Inc. & 10,958   \\ 
                N/S               & 9139   \\ 
                Let's Encrypt     & 7801    \\ \bottomrule
                \end{tabular}
            \caption{Certificate issuer distribution for SSTP servers. \label{tab:organizations-sstp}}}     
    \end{table}

Looking at the organization fields of the CA-signed certificates, we can learn more about the signing authorities. Considering SSTP, we filter out 2502 different organizations for almost 100,000 CA-signed certificates. \Cref{tab:organizations-sstp} contains the top five organizations accounting for 87\% of all signings. We examine the issuer CNs for the certificates that do not specify an organization, yet we could not find any meaningful information with 7,531 different issuers and the most frequently occurring CN being \textit{CA} with 159 signings. For OpenVPN, the organizations are a lot more heterogeneous with 14,548 organizations in total. The top five organizations in \Cref{tab:organizations-openvpn} account for only around 50\% of all signings.

    \begin{figure}
        \centering
        \subfloat[OpenVPN server certificates.\label{subfig:ecdf-openvpn}]{%
            \includegraphics[width=0.5\textwidth]{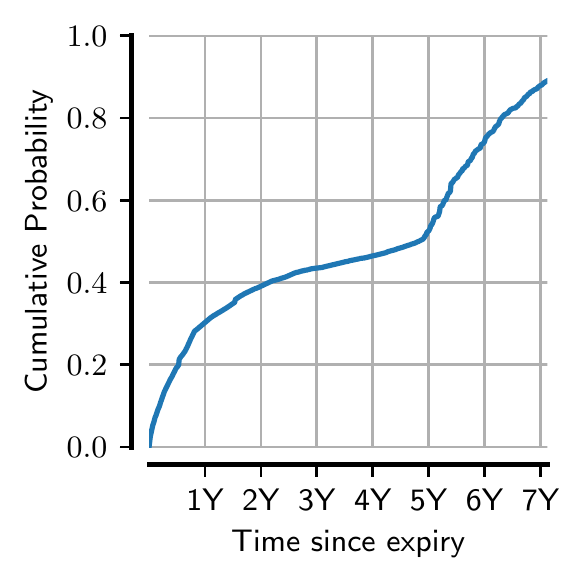}
        }
        \subfloat[SSTP server certificates.\label{subfig:ecdf-sstp}]{%
            \includegraphics[width=0.5\textwidth]{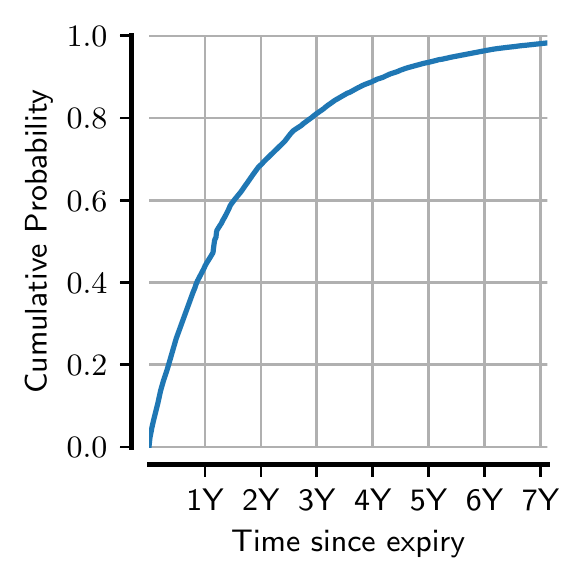}
        }
        \caption{Distribution of expiry time (time between day of expiry and Aug. 15, 2022) for expired certificates.}
        \label{fig:ecdfs}
    \end{figure}

Since we also detect a quite significant number of expired certificates, we examine the date of their expiry more thoroughly. \Cref{fig:ecdfs} shows ECDFs for the time that has passed since the dates of expiry and the 15th of August, 2022. In general, over half of the SSTP certificates expired over a year ago. For OpenVPN it is even around 70\%. It is possible that those certificates belong to outdated, forgotten VPN servers. 

\noindent\textbf{TLS vulnerability analysis}.
The results of our TLS vulnerability analysis for the TLS-based VPN protocols can be found in the last two columns of \Cref{tab:tls-vulnerabilities} where we count the occurrences of susceptible servers. We detect a larger number of vulnerable servers for RC4, Poodle and ROBOT for both protocols, yet only a few outliers for the rest. SSTP is much more likely to show signs of vulnerability for all three attacks with over 90\% of the servers being susceptible to ROBOT. This is most likely caused by the fact that SSTP is based on an outdated version of SSL and highlights why SSTP is not recommended to be used anymore. 
    
    \begin{table}
        \begin{adjustbox}{max width=\textwidth}
        \begin{tabular}{@{}llllll@{}}
        \toprule
                   & TLS version        & Cipher suites       & Other requirements & OpenVPN & SSTP    \\ \midrule
        RC4 \cite{alFardan2013}        & All            & RC4                & None               & 32,294  & 84,892  \\ 
        Heartbleed \cite{heartbleed} & All            & All                & OpenSSL Heartbeat  & 232     & 10      \\ 
        Poodle \cite{poodle}     & SSL 3.0        & All                & None               & 7,005   & 24,917  \\ 
        FREAK \cite{freak}      & All            & RSA\_EXPORT        & None               & 31      & 1       \\ 
        Logjam \cite{logjam}     & All            & DHE/512-bit export & None               & 8       & 0       \\ 
        DROWN \cite{drown}      & SSLv2          & All                & None               & 0       & 0       \\ 
        ROBOT \cite{robot}      & All            & TLS\_RSA           & None               & 95,301  & 174,986 \\ 
        Raccoon \cite{raccoon}    & TLS $\leq 1.2$ & TLS\_DH            & None               & 0       & 0       \\ \bottomrule
        \end{tabular}
        \end{adjustbox}
        \caption{Requirements for TLS vulnerabilities and number of vulnerable servers per protocol.}
        \label{tab:tls-vulnerabilities}
    \end{table}

\noindent\textbf{The effect of not using SNI}.
As we target only IP addresses in our follow-up TLS measurements without the SNI extension, we want to investigate the effect of not using SNI. Therefore, we first perform an rDNS resolution for our IP addresses and find 259,910 domain names for about 480,000 OpenVPN TCP servers and 86,630 domain names for roughly 180,000 SSTP servers. We now collect certificates with the SNI extension and then re-run the TLS scans without SNI for the respective addresses for whose domains we could gather certificates.

\Cref{tab:sni-comparison} shows the results of the comparison of those two types of certificates and the number of certificates we could collect. Two certificates mismatch when the fingerprints differ. We then compare different fields and summarize the mismatch occurrences in the table. If those fields match and the certificate has only been renewed, we do not count it as a mismatch. 

While the results for both protocols are similar, relatively speaking, we find more mismatches for SSTP. About 3\% mismatch for OpenVPN, whereas for SSTP 5.5\% mismatch. To confirm that those mismatches are caused by using SNI in the TLS handshakes, we perform a second measurement without SNI and compare the certificates with the other non-SNI results. Without SNI, we find less than half as many mismatches for SSTP and more than three times fewer mismatches for OpenVPN.

Considering the overall number of certificates from our large-scale measurements compared to the ones we collected with SNI and keeping in mind the mismatches we detected in the two non-SNI measurements, we can conclude that not using SNI affects less than 1\% of the certificates for both protocols and the effect is therefore negligible.

    \begin{table}
        \begin{tabular}{@{}lll@{}}
            \toprule
                                        & OpenVPN & SSTP   \\ \midrule
            SNI Certificates            & 84,212 & 45,405 \\
            no SNI Certificates         & 81,379 & 45,026 \\
            Certificate Mismatches      & 2491   & 2515  \\
            Authority Key ID Mismatches & 2051   & 1463  \\
            Subject Key ID Mismatches   & 2407   & 2379  \\
            Subject SANs Mismatches     & 2008   & 1677  \\
            Issuer CN Mismatches        & 1933   & 1476  \\
            Subject CN Mismatches       & 2021   & 1627  \\ \bottomrule
            \end{tabular}
        \caption{Comparison of Certificates Collected with and without SNI}
        \label{tab:sni-comparison}
        \end{table}

\subsection{Fingerprinting}
\budget{2.5}

\noindent\textbf{Server Software}. 
For SSTP and PPTP, we can infer the server-side software from the responses we receive to our initiation requests. For SSTP, we find that around 80\% of all detected servers use \textit{Microsoft HTTPAPI 2.0}. Around 19\% use \textit{MikroTik-SSTP} and less than 1\% use something else or specify nothing at all.

However, the PPTP vendor software is a lot more heterogeneous compared to SSTP. \Cref{tab:pptp-vendors} shows the different software vendors we detect in the VPN server responses. While there are four prominent vendors, over 15\% of the PPTP servers rely on 183 different types. This can have potential security implications on the PPTP ecosystem. Assuming there was some kind of new vulnerability, the rollout of a security update to counter this vulnerability would be significantly slower compared to SSTP with fewer software vendors. \new{A similar phenomenon where vendor fragmentation leads to slower update rollout can also be observed in the Android ecosystem. Thomas et al. \cite{thomas2015} showed that almost 60\% of all devices ran insecure Android versions in July 2015. This share declines only slowly after the discovery of a major vulnerability. They found out that the bottleneck of this issue lies with the manufacturers and results in 87.7\% of all devices being exposed to at least 11 critical vulnerabilities. Jones et al. \cite{jones20} considered manufacturers between 2015 and 2019 and further showed that the median latency of a security update is 24 days with an additional latency of 11 days before an end-user update.}

    \begin{table}
        \begin{tabular}{lr}
        \toprule
        Vendor              & Percentage  \\ \midrule
        Linux               & 32.3\% \\ 
        MikroTik            & 30.6\% \\ 
        Draytek             & 21.1\% \\ 
        Microsoft           & 6.9\%  \\ 
        Cananian            & 2.0\%  \\ 
        Fortinet PPTP       & 1.4\%  \\ 
        Yamaha Corporation  & 1.4\%  \\ 
        Cisco Systems, Inc. & 1.2\%  \\ \midrule
        Others (162)          & 3.2\%  \\ \bottomrule
        \end{tabular}
        \caption{Software vendors for detected PPTP servers.}
        \label{tab:pptp-vendors}
    \end{table}

\begin{figure}[t!]
    \begin{center}
    \includegraphics[width=0.8\linewidth]{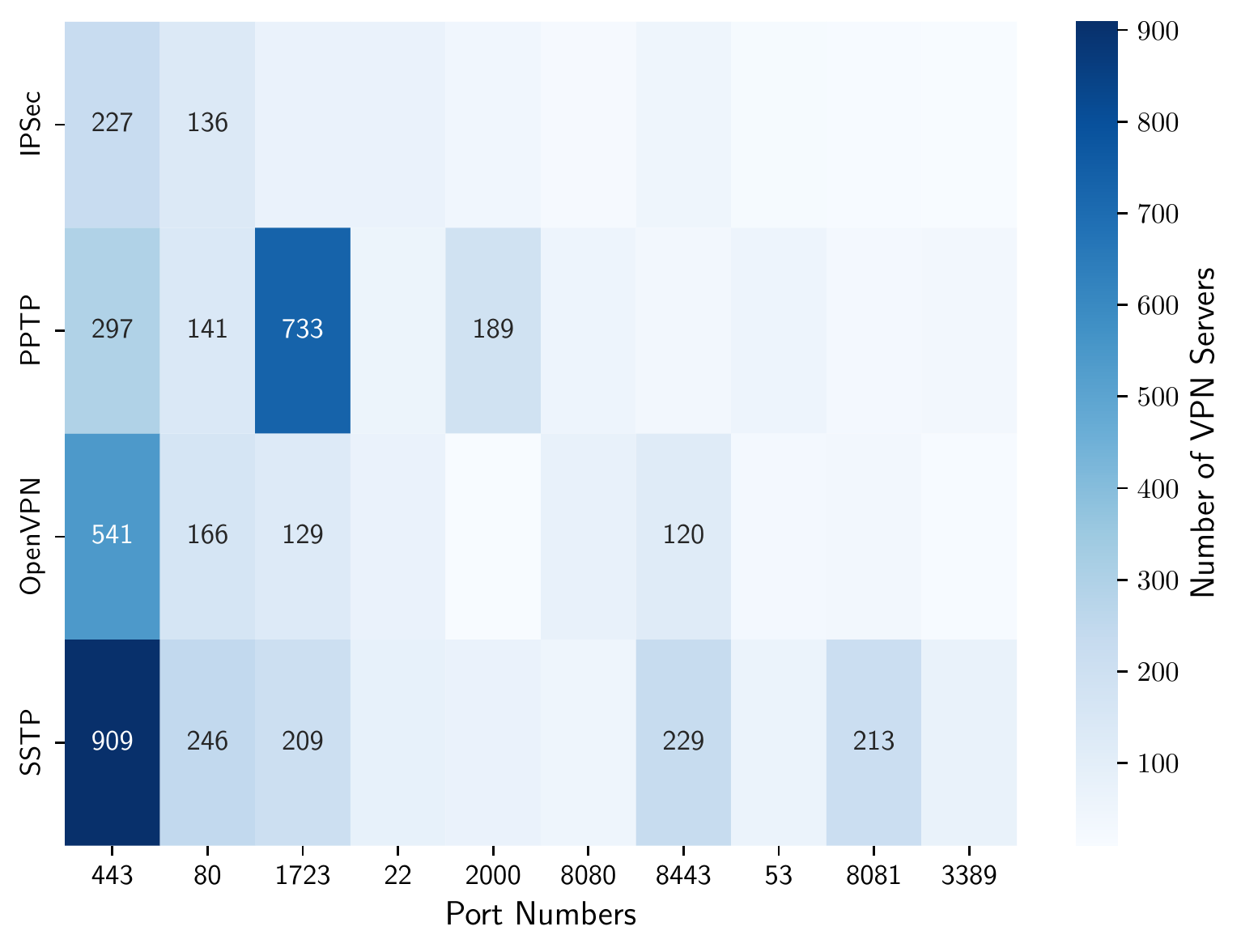}
    \end{center}
    \caption{Heatmap of most frequently detected open ports per VPN server.}
    \label{fig:port-heatmap}
\end{figure}

\noindent\textbf{Nmap OS detection and port scans}.
In our Nmap OS detection measurements, we first have a look at the most common ports for all four protocols. \Cref{fig:port-heatmap} summarizes the most frequently occurring open ports. As expected, the default HTTP(S) ports 443 and 80 are the most common ports, with the exception of the PPTP servers for which the default PPTP port TCP/1723 obviously is the most widely used port. As Ramesh et al. \cite{ramesh2022} pointed out, specific open ports do not pose security risks by themselves, yet, they might still be abused in order to identify and exploit particular services \cite{tcp-ports-to-test}.

    For the OS detection, we filter out the first guesses for every target and look at the most common OSes and version ranges:
    \begin{itemize}
        \item \textbf{IPsec:} We receive 48 unique first guesses for 126 hosts out of 722 responsive IPsec servers. Out of those, 40 guess the Linux Kernel ranging from version 2.6.32-3.10. In general, Linux is the most common OS with 67 guesses. However, Microsoft was barely guessed as an OS vendor with only nine guesses.
        \item \textbf{PPTP:} For 792 responsive hosts, Nmap was able to guess an OS for 216 addresses with 56 unique guesses. Linux was once again the primary occurrence. Out of those guesses, 88 specified Linux 2.6.32-3.10, where the majority mew{lie} below version 3.2, however. As for IPsec, we have very few results for Microsoft with only 15 guesses. For the PPTP servers, there were more hardware guesses compared to the other protocols with 36 guesses specifying some kind of hardware device.
        \item \textbf{OpenVPN:} The most frequent guesses are almost exclusively Linux again in 33 unique guesses for 89 out of the 763 responsive hosts. 39 specify Linux ranging from 3.2-4.11, \ie the versions are not quite as outdated as for PPTP and IPsec. We received only a single guess for Microsoft products.
        \item \textbf{SSTP:} The SSTP scans result in 44 different guesses for 178 out of 948 responsive hosts. This time, we have more results for Microsoft products with a total of 49 guesses. The most prominent vendor is Linux again, however, with 101 guesses where 53 range from Linux versions 2.6.32-3.10.
    \end{itemize}

\subsection{IPv6}
\budget{2.5}

\noindent\textbf{VPN server detection}.   
Targeting roughly 530 million IPv6 addresses in our \zmapsix port scans, we could detect 1,195,510 responsive hosts on port TCP/443 which we target in our follow-up \zgrab scans for SSTP and OpenVPN over TCP. We could not find any responsive addresses on port TCP/1723, the default PPTP port. Since port TCP/1723 is used exclusively for PPTP and the protocol is very outdated, it is not too surprising that there are no IPv6 servers supporting PPTP. Apart from that, we do not get any responses on the UDP ports 500 (IPsec) and 1194 (OpenVPN over UDP). 

Out of the roughly 1.2 million hits on port TCP/443, we could identify 2070 addresses as OpenVPN servers and 949 as SSTP servers with a total of 2221 VPN servers supporting IPv6. While those results seem very low, we have to keep in mind that the rollout of IPv6 is still very slow in general. IPv6 is also not yet supported by most commercial VPN providers.

As also observed in IPv4 results in \Cref{active:vpn-protocols}, none of the OpenVPN servers accepted our OpenVPN key method 1 requests with only 11 servers still responding with the secure key exchange method. Additionally, of the overall IPv6 VPN servers we detect, around 36\% support both protocols, \ie compared to IPv4, the overlap is higher.

\new{Investigating the rDNS records corresponding to the responsive IPv6 addresses, we observe that the top 10 domains belong to hosting providers, cloud providers, and research networks. 
Similar to the IPv4 results, we do not find a domain name belonging to a commercial VPN provider among the top 10 domains.
By filtering second-level domains to match \textit{*vpn*} we find the commercial VPN provider WhiteLabel VPN, ranking 25th among the top domains.}

\new{Therefore, we infer that most of the VPN servers that support IPv6 are, in fact, not commercial VPN providers.}

\noindent\textbf{TLS certificate analysis}.
The results of the TLS certificate analysis are similar to IPv4. We could collect certificates for around 75\% of the identified OpenVPN servers with 816 unique fingerprints. Combined with the certificates that do not contain a fingerprint, we analyze a total of 1882 certificates. We collected certificates for every SSTP server resulting in 747 certificates after filtering out 207 unique fingerprints. Less certificates are expired this time with only 3.3\% for OpenVPN and 2.1\% for SSTP. This time, only 29\% of the OpenVPN certificates are self-signed. For SSTP, more certificates are self-signed for the IPv6 servers with over 70\% of all certificates. Out of those, we characterize roughly 2\% as snake oil certificates for both protocols. Furthermore, about two thirds of the self-signed certificates for both protocols were issued by softether.

When examining the signing organizations for the CA-signed certificates, we find that around 85\% (709 certificates) of the OpenVPN certificates are signed by Let's Encrypt with a total of 43 organizations. For SSTP, around 73\% are signed by Let's Encrypt (153 certificates). Here, we find a total of only 16 organizations.

\noindent\textbf{TLS vulnerability analysis}. 
The results of the TLS vulnerability analysis are very similar to the IPv4 VPN servers. For both protocols, we are only able to detect vulnerable servers for the same three prominent attacks as for the IPv4 analysis. Out of the 2070 OpenVPN servers, 31\% are vulnerable to RC4 biases, 6\% to Poodle and 74\% to Robot. When analyzing the 949 SSTP servers, we find that  67\% are vulnerable to RC4 biases, 13\% to the Poodle attack and roughly 98\% to ROBOT. While the results are similar to our large-scale measurements, we can conclude that the VPN servers supporting IPv6 are much more likely to show any signs of vulnerability with the vast majority being vulnerable to the ROBOT attack.

\noindent\textbf{The effect of not using SNI}.
The rDNS measurements for the IPv6 servers resulted in 410 domain names for SSTP and 813 domain names for OpenVPN over TCP. Again, we first collect TLS certificates using the SNI extension and then try the same without SNI and compare the results. We find that only around 3\% of the certificates for both protocols mismatch in terms of fingerprints and important certificate fields including authority and subject key IDs, subject SANs, and CNs. When comparing those results by running a second TLS scan without SNI, we find that only around 2.5\% of the OpenVPN and less than 1\% of the SSTP certificates differ. Considering the overall number of certificates, the effect of not using SNI is even less significant compared to IPv4 and is therefore negligible. 

\noindent\textbf{VPN server software}.
Since we could not analyze the PPTP server software ecosystem this time, we can only compare the results for SSTP. The results are similar again with 91\% of the SSTP servers specifying the Microsoft HTTP API 2.0. However, the rest did not specify any vendor, \ie the IPv6 SSTP servers seem to not use MikroTik-SSTP with Microsoft being the only vendor. 

\noindent\textbf{Nmap OS detection and port scans}.
As for IPv4, we perform Nmap measurements on the detected IPv6 VPN servers including 1000 random OpenVPN TCP servers and all 949 SSTP servers. Out of those servers, 874 OpenVPN servers and 852 SSTP servers are responsive.

The most commonly used open port is TCP/443 with 838 occurrences (96\%) for OpenVPN and 852 (97\%) for SSTP. Compared to IPv4, the number of open HTTPS ports is much higher for OpenVPN. Here, we have to keep in mind that we can only consider OpenVPN servers over TCP. Thus, this disparity is to be expected. The second most frequently open port for both protocols, in contrast to IPv4, is TCP/22, the default SSH port. This port occurs 245 times (28\%) for OpenVPN and even 391 times (46\%) for SSTP. Other common ports for both protocols are ports TCP/8000 for OpenVPN (21\%) and TCP/80 accounting for around 17\% of the open ports for both protocols.

We receive more OS guesses for the IPv6 servers compared to IPv4. As was the case for IPv4, we filter out the first guesses for every target:

\begin{itemize}
    \item \textbf{OpenVPN}: The measurement results in only four unique guesses for a total of 481 hosts. 93\% specify Linux with 416 guessing Linux 3.X and 33 guessing version 2.6. Only 19 predictions include a Microsoft OS and only 13 an Apple product.
    \item \textbf{SSTP}: For SSTP, there are five unique predictions for 406 addresses. The majority specifies Linux again with 91\%. Out of those, 333 guesses specify Linux version 3.X and only 36 specify version 2.6. Microsoft OSes are predicted 36 times and only a single guess specifies a macOS. 
\end{itemize}

In contrast to IPv4, Nmap was able to predict an OS for a much larger percentage of our targets with an OS guess for almost half of the targets. Additionally, the predictions are a lot more homogeneous. Linux is again the most prominent vendor, however, the predicted versions are not quite as outdated as for the IPv4 servers.
    \section{Passive VPN Traffic Analysis}
\budget{2}
It is important for network operators and ISPs to gain insight over the volume and daily patterns of VPN traffic.
In a previous study, Feldmann et al. \cite{Feldmann2020LockdownEffect} try to find the VPN traffic based on the domain names corresponding to the IP addresses observed in the traffic.
For detecting VPN traffic, Feldmann et al. use domain names to infer whether the IP addresses corresponding to them carry VPN traffic.
They exclude any domain name that starts with \textit{www.}, and does not have \textit{*vpn*} to the left of the public suffix.
Finally, they consider the remaining domain names as VPN domain names and count the traffic that relate to these domain names as VPN traffic.

To compare our methodology with the state of the art, we apply the methodology used by Feldmann et al. \cite{Feldmann2020LockdownEffect} on our results. 
We use DNS responses gathered by DNS resolvers at a large European ISP, and look for those DNS responses that include the IP addresses from our VPN hitlist. 
We find 13\% of the IP addresses from the VPN hitlist in the above-mentioned DNS responses. 
Therefore, we complement our DNS data with reverse DNS look-ups for all the remaining IP addresses.
To refine the reverse DNS results, we exclude any domain names containing any order of the corresponding IP address bytes or octets in decimal or hexadecimal format. 
Overall, we end up with the domain names corresponding to 23.6\% of the IP addresses from the VPN hitlist.
Then, we apply the methodology used by Feldmann et al. on the resulting domain names, i.e., we extract those domain names that contain \textit{*vpn*} on the left side of the public suffix \cite{public-suffix}, while excluding any domain starting with \text{www.} to exclude web servers.
We observe that this methodology captures only 4.8\% of our VPN hitlist. Therefore, our approach can detect 4 times more VPN servers compared to the methodology by Feldmann et al.

Finally, we look at \new{a one-week snapshot of all the} network flow traffic from the large European ISP to find out the amount of traffic that can be attributed to VPN.

To this end, we compare the amount of VPN traffic detected with three methodologies:
\begin{enumerate}
\item \textit{VPN Hitlist}: the methodology proposed in this paper, i.e. sending active probes, including the responsive IP addresses in a hitlist, excluding those IP addresses that answer to web requests, i.e. HTTP GET requests, then measuring the traffic volume originated by or destined to these IP addresses.
\item \textit{Port-based}: this methodology captures the traffic only based on port numbers, considering traffic with port numbers 500 (IPsec), 4500 (IPsec), 1194 (OpenVPN), 1701 (L2TP), 1723 (1723) both on UDP and TCP as VPN traffic.
\item \textit{Domain-based}: the methodology proposed by Feldmann et al, i.e. filtering domain names based on certain keywords, then measuring the traffic volume originated or destined to the IP addresses corresponding to these domain names.
\end{enumerate}

\begin{figure}
  \begin{center}
  \includegraphics[width=\linewidth]{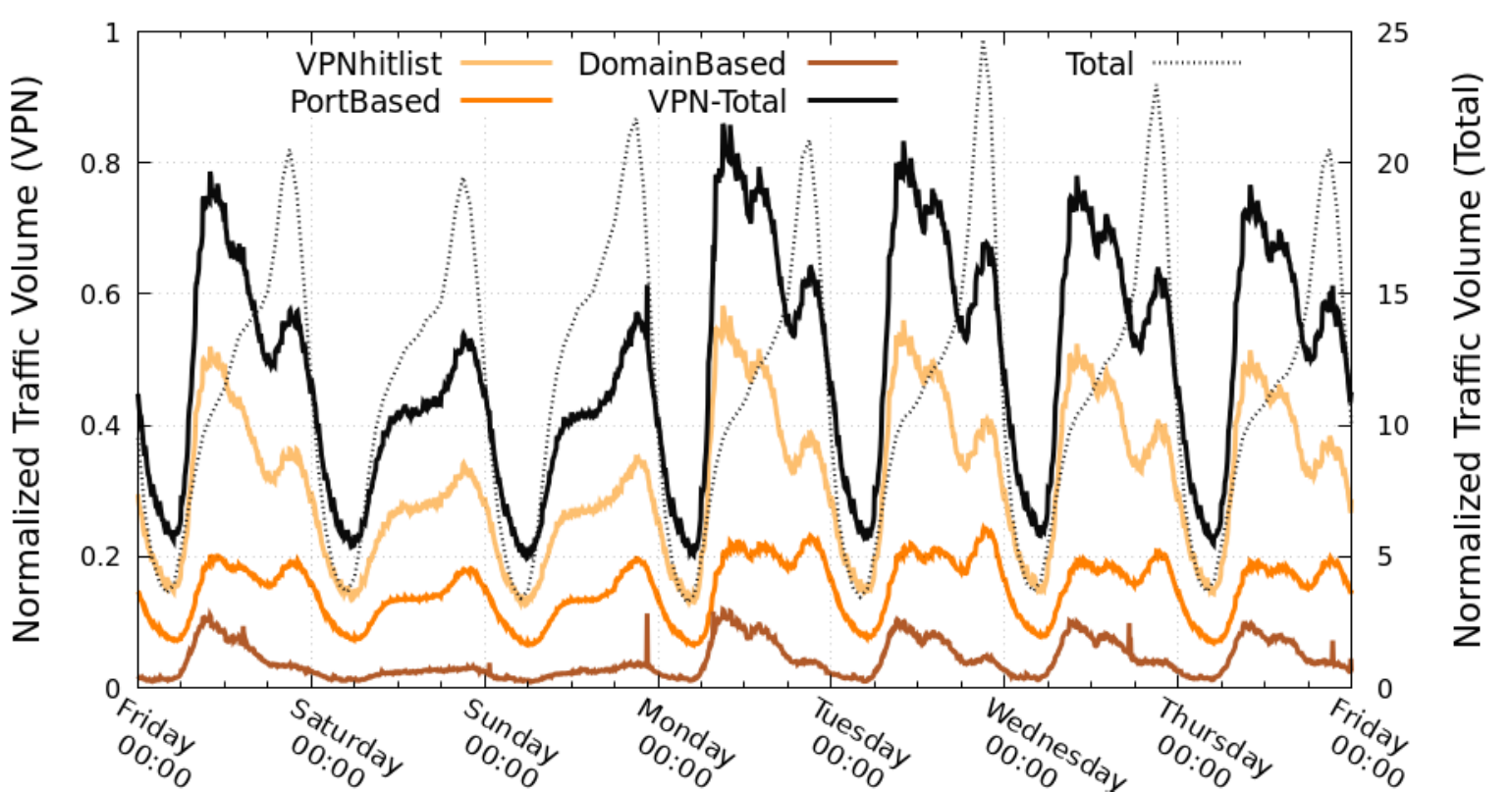}
  \end{center}
  \caption{Normalized VPN traffic volume for different traffic detection techniques.}
  \label{fig:vpn-traffic}
\end{figure}
 
\Cref{fig:vpn-traffic} shows the traffic volume considered as VPN traffic by each of the above-mentioned methodologies. 
The solid black line shows the total amount of VPN traffic detected by either of the three approaches. The dashed line shows the total traffic volume in the ISP.
The left Y axis shows the VPN traffic volume (including all the three approaches), and the right Y axis shows the total ISP traffic volume. All the traffic values are normalized. 
\new{While normalizing, we keep the ratio between the VPN traffic and total traffic intact. Therefore, comparing the left and right axis values shows that the total traffic is roughly 25 times as much as all VPN traffic.}

Compared to the \textit{Port-based} approach, we detect twice as much traffic, and compared to the \textit{Domain-based} approach, we detect 8 times as much using the \textit{VPN Hitlist}.

The mean VPN traffic volume detected by all three approaches is 4.1\% of the mean total ISP traffic over the week, with \textit{VPN Hitlist} contributing to 2.6\%, \textit{Port-based} 1.3\%, and \textit{Domain-based} 0.3\%. 

Looking at the overlap between every two approaches, we find that only 2.7\% of all the traffic detected by all three approaches is detected both by \textit{VPN Hitlist} and \textit{Domain-based}. We observe 1.2\% overlap between the traffic detected by \textit{VPN Hitlist} and \textit{Port-based} approaches.

\new{We observe a diurnal pattern in the VPN traffic detected by all of the three approaches.}
We find that VPN traffic pattern in the weekdays differs from that of weekends. It peaks at noon in the weekdays, and at night in the weekend, while the total ISP traffic always follows the same pattern, i.e. peaks at night.
It could indicate the fact that the VPN traffic is mostly work-related through weekdays, while mostly entertainment-related throughout the weekend.
In the domain-based approach the amount of VPN traffic detected by the \textit{Domain-based} approach is much less in the weekends than in the weekdays. This could indicate that the \textit{Domain-based} approach detect mostly work-related VPN servers.

We investigate the domain names corresponding to the traffic we detect using our approach and find that \textit{vpn.}, \textit{mail.}, \textit{www.}, and \textit{remote.} are among the most common prefixes left to the public suffix part of the domain names, with \textit{vpn.} being the most common prefix.
The fact that we observe \textit{mail.} and \textit{www.} might be either re-use of the same domain name for other purposes by the network operators, or a mislabeling effect from our approach caused by not answering our HTTP Get requests.
\new{Also, looking at the DNS records corresponding to the IP addresses from our hitlist, using FlowDNS---a system to correlate DNS and Netflow data at scale \cite{flowdns}---we find that 5 out of 10 top domains are related to commercial VPN providers and the rest are CDN domains.}
We observe that the most common source port/destination port combination is 4500/4500 which belongs to IPsec, also port number 1194 which is registered for OpenVPN, and at the same time 1193, which is practically used for VPN \cite{openvpn1193}. We also observe that 51820/51820 and 1337/1337 which belongs to WireGuard protocol are among the top port number pairs observed in the traffic detected by our approach. \new{Port 51820 also falls into the range of ephemeral ports numbers (49152 to 65535) which can be temporarily used by many applications. However, due to the prominent existence of this port number in our results accompanied with port 1337, we infer that we can possibly detect some WireGuard traffic, although in our active measurement approach we cannot scan the WireGuard protocol. This might be due to the co-existence of multiple protocols in one VPN server.}
This shows that although in our approach we cannot scan WireGuard protocol, we can still detect some WireGuard traffic which might be due to the co-existence of multiple protocols on one VPN server. 
Traffic related to WireGuard protocol contributes to 8.6\% of the detected VPN traffic by our VPN hitlist, while contributing to only 2\% of the traffic detected by the \textit{Domain-based} approach.
\section{Discussion}
\budget{1.5}
In this work, we detect VPN servers in the wild by sending Internet-wide active probes using different VPN protocols. 
We can distinguish between VPN servers and Web servers by excluding those servers that respond to a Web request.
We compare the amount of traffic detected by our approach and two other approaches over a week of traffic from a large European ISP and find out that the approach proposed by this work detects much more VPN servers compared to the state-of-the-art domain-based approach. 
In addition, our approach benefits from detecting VPN servers that do not use any domain name, and can also detect VPN traffic that is using unusual ports in case these servers answer VPN probe on the usual VPN port numbers.
Also, to be the best of our knowledge, this is the first work to perform an Internet-wide active measurement of VPN servers in the wild.

\noindent\textbf{\new{VPN hitlist.}}
We send active probes according to the specification of VPN protocols including SSTP, PPTP, OpenVPN, and IPsec to the whole IPv4 address space and to an IPv6 hitlist. 
We make our list of detected VPN servers, namely the VPN hitlist, publicly available at \href{https://vpnecosystem.github.io/}{vpnecosystem.github.io}.
This VPN hitlist can be useful for network operators to find out about the amount and patterns of VPN traffic in their networks.
The VPN hitlist can also be used by fellow researchers to investigate different behaviors of the VPN servers and VPN traffic, e.g. investigating actual attacks to these servers.

\noindent\textbf{\new{Security.}}
We also investigate the security of the OpenVPN and SSTP protocols in terms of different security aspects, including heartbleed attack, TLS certificates security, and TLS downgrade attacks.
We find that SSTP servers use expired certificates 3x more than OpenVPN servers. 
We also find that 90\% of the SSTP servers are vulnerable to ROBOT attack. 
Therefore, we find SSTP to be the most vulnerable protocol. 
This striking high percentage of vulnerable servers for some of the protocols shows, that the VPN \new{server} ecosystem is not as secure as some users believe it to be.
Therefore, we hope that our analysis can highlight these security risks with using each VPN protocol and also helps network operators choose the right VPN protocols for their networks.

\noindent\textbf{\new{Limitations. }}
Our approach builds upon receiving answers from the servers in the wild and therefore, has its limitations.
If there is a VPN protocol which uses a pre-shared key in the first VPN request and does not respond otherwise, we are unable to detect it. Examples of such VPN protocols are WireGuard and Cisco AnyConnect. Therefore, we are unable to detect any VPN server which offers only these two protocols. 
However, we observe that 8.6\% of the detected traffic is related to WireGuard which might be due to multiple protocols being served by one VPN server. 
In addition, certain VPN servers might only work on non-registered port numbers for better anonymization. Since in our work, we only send probes to the port numbers registered for the VPN protocols by IANA \cite{iana}, we cannot detect VPN servers that work on unusual port numbers.
\new{Therefore, our list of detected VPN servers is limited to those using the supported VPN protocols and working on their registered port numbers.}
 
\noindent\textbf{\new{Future work.}}
In the future, our work can be complemented by including more port numbers in the active scans. Results from previous studies on predicting the services across all ports \cite{predictports2022} can be used together with our approach to gain more coverage.
Despite the above-mentioned limitations, our proposed approach detects much more VPN servers compared to the state-of-the-art domain-based approach, and also, to the best of our knowledge, is the first work to perform an Internet-wide active measurement of VPN servers in the wild.

\noindent\textbf{\new{Reproducibility.}}
We make our analysis code \new{and data \cite{edmond-dataset}}, customized \zgrab modules \cite{zgrab2-vpn}, and our VPN hitlist publicly available\footnote{\new{\url{https://vpnecosystem.github.io/}}} for fellow researchers to be able to reproduce our work and build upon it. 

\begin{comment}
\begin{itemize}
    \item What are the implications of our findings for researchers, network operators, security,\dots?
    \item How do they change the VPN ecosystem? Which protocols are doing well? How would it evolve in the future?
    \item What are the limitations of our work?
\end{itemize}
\end{comment}
 %
\section{Related Work}
\budget{1.5}

VPN traffic classification is an open research problem, particularly challenging due to its encrypted nature. 
There are several studies trying to tackle this problem using machine learning approaches. 
Some are able to categorize the traffic into VPN and non-VPN only \cite{lotfollahi2020deep}, and some provide more detailed sub-categories \cite{8622812,8004872,alfayoumi2022}.
Zou et al. \cite{8622812} identify encrypted traffic by combining a deep neural network to extract features of single packets and a recurrent network to analyze features of the traffic flow based on features of three consecutive packets. 
Though the model classifies some traffic incorrectly regarding sub-categories, it could achieve almost 99\% accuracy when only considering VPN
and non-VPN traffic. Alfayoumi et al. \cite{alfayoumi2022}, on the other hand, also consider time-related features and subdivide traffic by also identifying applications.

All of these works require previously captured unencrypted VPN traffic to train.
Previous studies have also tried to detect VPN traffic using the DNS records corresponding to the IP addresses observed in the traffic \cite{Feldmann2020LockdownEffect}. 

In this paper, we propose a different approach, i.e. Internet-wide active measurements, to detect VPN servers in the Internet. 
Internet-wide measurements have been previously applied for several intents including finding IPv6 responsive addresses \cite{gasser2018clusters}, responsive IPs to abnormal traffic \cite{port0}, the usage of DNS over encryption \cite{dotactive}, and so on. 
However, to the best of our knowledge, this is the first work applying active measurements to detect VPN servers in the wild and detecting the traffic based on a VPN hitlist.

Investigating the security of the VPN servers is also an interesting research problem which is already addressed by several studies. For example, Xue et al. investigate the possibility and practicality of fingerprinting OpenVPN flows \cite{280012}. Tolley et al. investigate the vulnerability of known VPN servers to spoofed traffic \cite{272192}. 
Crawshaw \cite{Crawshaw2020} addresses vulnerabilities that come with some of the protocols themselves, such as outdated cryptographic cipher suites used in PPTP. In his proposal for WireGuard \cite{donenfeld}, Donenfeld talks about disadvantages in current popular VPN protocols.
\textit{VPNalyzer} requires a tool to be installed on the user's device to measure and collect data on the active VPN connections in terms different security aspects including data leakage, open ports, and DNSSEC validation \cite{ramesh2022}.
Appelbaum et al. also identified vulnerabilities of commercial and public online VPN servers \cite{appelbaum2012vpwns}.

A large body of literature also exists that empirically examines TLS vulnerabilities including self-signed root CA injection to intercept TLS connection \cite{raman20, ikram2016}, and improper implementation of the protocol making version downgrade attacks possible even with new TLS 1.3 \cite{lee20}.

We mainly focus on potential vulnerabilities that come with VPN protocols which are built on top of SSL/TLS. Thus, we investigate SSL/TLS related features of those protocols. For some identified OpenVPN servers, we can also make assumptions on their security based on information we can infer about their server configurations.
All the previous works study the security of known VPN servers, while in this paper, we measure the vulnerability of our detected VPN server in the Internet.
\section{Conclusion}
\budget{0.5}
In this paper, we performed the first Internet-wide active measurement on the VPN \new{server} ecosystem for OpenVPN, SSTP, PPTP, and IPsec both in IPv4 and IPv6 to detect VPN servers in the wild.
We detected 9.8 million VPN servers distributed globally. 10\% of the detected VPN servers offered more than one VPN protocol with very few serving all the four protocols we studied. 
We also send active Web probes to the detected VPN servers and observed that 2\% were both VPN and Web servers. Analyzing the TLS-based VPN protocols, i.e. OpenVPN and SSTP, we found that SSTP was the most vulnerable to a version downgrade attack, and certificates of OpenVPN servers had the most self-signed and self-issued certificates. We also tried to fingerprint the detected VPN servers in terms of server software vendors and operating systems.
Finally, using our VPN hitlist, excluding the servers that were both VPN and Web servers, we observed that VPN traffic constitutes 2.6\% of the total traffic volume in a large European ISP, which is 8x as much as that of a state-of-the-art domain-based approach, and twice as much as the trivial port-based approach.
We publish our VPN hitlist, our customized ZGrab2 modules for VPN scans, and the code to our analysis for future researchers and network operators to use.
\balance

\bibliographystyle{splncs04}
\bibliography{paper}

\end{document}